\documentclass[english,superscriptaddress,prl,preprint]{revtex4-1}
\usepackage{lmodern}

\usepackage[T1]{fontenc}
\usepackage[latin9]{inputenc}
\usepackage{color,soul}
\usepackage{amsmath}
\usepackage{amssymb}
\usepackage{graphicx}

\makeatletter

\usepackage{float}

\DeclareMathOperator{\rank}{rank}

\usepackage{babel}

\makeatother

\usepackage{babel}
\begin{document}
\title{Detecting Hidden Units and Network Size\\ from Perceptible Dynamics}
\author{Hauke Haehne}
\email{hauke.haehne@uni-oldenburg.de}

\affiliation{Institute of Physics and ForWind, University of Oldenburg, 26111 Oldenburg,
Germany}
\author{Jose Casadiego}
\affiliation{Chair for Network Dynamics, Institute for Theoretical Physics and
Center for Advancing Electronics Dresden (cfaed), Technical University
of Dresden, 01062 Dresden, Germany}
\author{Joachim Peinke}
\affiliation{Institute of Physics and ForWind, University of Oldenburg, 26111 Oldenburg,
Germany}
\author{Marc Timme}
\email{marc.timme@tu-dresden.de}

\affiliation{Chair for Network Dynamics, Institute for Theoretical Physics and
Center for Advancing Electronics Dresden (cfaed), Technical University
of Dresden, 01062 Dresden, Germany}

\begin{abstract}
The number of units of a network dynamical system, its size, arguably
constitutes its most fundamental property. Many units of a network,
however, are typically experimentally inaccessible such that the network
size is often unknown. Here we introduce a \emph{detection matrix
}that suitably arranges multiple transient time series from the subset
of accessible units to detect network size via matching rank constraints.
The proposed method is model-free, applicable across system types
and interaction topologies and applies to non-stationary dynamics
near fixed points, as well as periodic and chaotic collective motion.
Even if only a small minority of units is perceptible and for systems
simultaneously exhibiting nonlinearities, heterogeneities and noise,
\emph{exact} size detection is feasible. We illustrate applicability
for a paradigmatic class of biochemical reaction networks. 
\end{abstract}

\maketitle

\begin{figure}[t]
\includegraphics{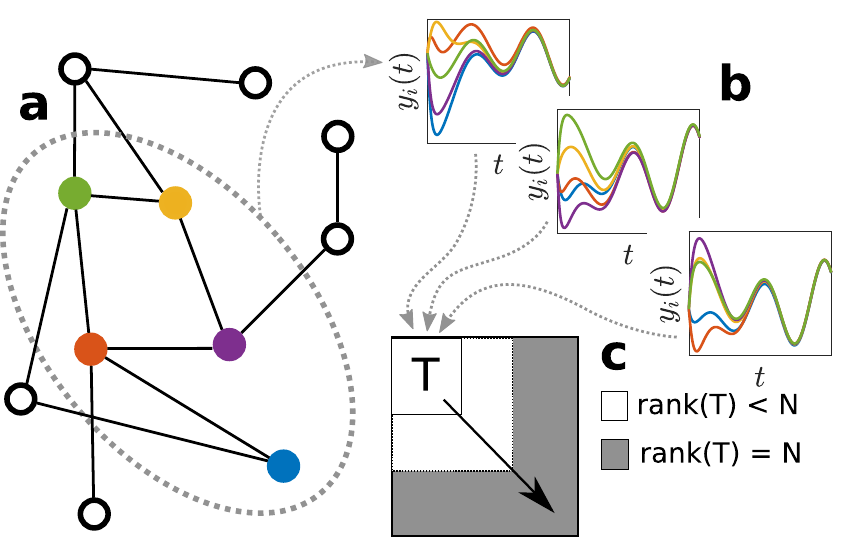} \caption{\textbf{Revealing network size from the dynamics of perceptible units.}
\textit{(a)} Scheme of a network of $N$ units where only $n<N$ units
(colored disks, encircled by dashed line) are accessible for measurement
(perceptible). \textit{(b)} Transient time series measured from accessible
units, started from different initial conditions (trajectory colors
match observable units in (a)). \textit{(c)} Observed nonlinear, multi-dimensional
time series are arranged into the \emph{detection matrix} $T_{k}$,
satisfying the condition $\rank\left(T_{k}\right)=N$ if and only
if $kn>N$ and $M>N$, according to \eqref{eq:CORE} introduced below.}
\label{fig:sketch} 
\end{figure}

Networks of interacting dynamical units prevail across natural and
human-made systems \cite{Strogatz2001a,Newm10,motter2018antagonistic}.
Examples range from intra-cellular gene-regulatory networks critical
for survival \cite{Karlebach2008,Huang2017} to power grids supplying
electric energy on demand \cite{filatrella2008analysis,rohden2012self,menck2013basin,motter2013spontaneous,menck2014dead,Witthaut2016}
and to social and transportation networks determining how ideas and
diseases spread \cite{brockmann2013hidden,Sun2014,Horvath2014}. Key
properties of the physical interaction topology in such networks fundamentally
underlie their function such that revealing them from measurements
of the collective network dynamics constitutes a topical field of
research \cite{Yeung2002,Gardner2003,Yu2006,Timme2007,Yu2011,Shandilya2011,barzel2013network,timme2014revealing,Han2015a,casadiego2015network,Mangan2016,Casadiego2017,Nitzan2017}.

However, dynamical data from many networks are often only incompletely
accessible, because many of their units are hidden from measurements.
Thus the dynamics of a possibly small subset of units might be available
only. Such hidden units typically complicate the inference of direct
interactions by correlating or decorrelating the dynamics of measured
units in unpredictable ways \cite{Soudry2015,lunsmann2017transition}.
Nevertheless, partial information about a networked system may provide
hints about overall features of the network. For instance, approximating
the network dynamics via model differential equations may help to
detect the existence and location of a single hidden unit through
heuristics performed on reconstructed connectivity matrices for different
time windows \cite{Su2012,Shen2014,Su2016}. Other schemes exploit
dynamics to determine paths from observed, via hidden, to observed
units \cite{goncalves2007dynamical,goncalves2008necessary,Yuan11}
and typically require to know the exact number of hidden units \emph{a
priori}. Yet, how to reveal the number of many hidden units, or equivalently,
the overall network size from time series recorded from the collective
dynamics of accessible units remains generally unknown.

Here, we show that measuring the transient collective dynamics of
a subset of perceptible network units (accessible to measurement)
may robustly reveal the exact number of hidden units and thus identify
the network size. We demonstrate how specifically grouping different
transient time series obtained from perceptible units into a \emph{detection}
\emph{matrix} yields bounds relating the rank of such matrix to the
size of the full network, see Fig.~\ref{fig:sketch}. We propose
a simple detection algorithm to exactly find the number of hidden
units, even if they are the minority by far. The number of time series
necessary to reliably identify network size only linearly scales with
network size, thus making size detection scalable. The proposed method
generalizes from linear and linearized dynamics near fixed points
to dynamics near periodic orbits as well as to collective irregular
and chaotic dynamics, without requiring knowledge of a system model.
Even for systems simultaneously exhibiting nonlinearities, heterogeneities,
and noise detection may be feasible and exact.

\emph{Theory of detecting network size from observed dynamics. }Consider
a network dynamical system 
\begin{equation}
\dot{\boldsymbol{z}}=\boldsymbol{F}(\boldsymbol{z}),\label{eq:nonlinear}
\end{equation}
of an unknown number $N$ of coupled units $i\in\{1,\ldots,N\}$ where
$\boldsymbol{z}(t):=[z(t),z_{2}(t),$ $\ldots,z_{N}(t)]^{\mathsf{T}}\in\mathbb{R}^{N}$
is the system's state at time $t\in\mathbb{R}$ and $\boldsymbol{F}:\,\mathbb{R}^{N}\rightarrow\mathbb{R}^{N}$
an unknown smooth function that defines its rate of change and thereby
the collective network dynamics. For simplicity, we first present
the idea of identifying network size for noise-free linear dynamics
close to fixed points and below discuss how it generalizes to more
complex dynamics, including periodic and aperiodic, irregular dynamics,
e.g., noisy and collective chaotic motion. Close to a fixed point
$\boldsymbol{z}^{*}$ where $\boldsymbol{F}(\boldsymbol{z}^{*})=0$,
a first order approximation of \eqref{eq:nonlinear} in terms of $\boldsymbol{x}(t)=\boldsymbol{z}(t)-\boldsymbol{z}^{*}$
yields 
\begin{equation}
\boldsymbol{\dot{x}}(t)=A\boldsymbol{x}(t)\label{eq:linear}
\end{equation}
where $A\in\mathbb{R}^{N\times N}$ with elements $A_{ij}=\partial F_{i}/\partial x_{j}\left(\boldsymbol{z}^{*}\right)$
is the Jacobian matrix of $\boldsymbol{F}$ evaluated at $\boldsymbol{z}^{*}$
and defines an unknown proxy for the connectivity of the system, i.e.
$A_{ij}\neq0$ if unit $j$ directly acts on $i$ and $A_{ij}=0$
otherwise. Solving \eqref{eq:linear} yields $\boldsymbol{x}(t)=\exp(At)\boldsymbol{x}(0),$
where $\boldsymbol{x}(0)\in\mathbb{R}^{N}$ is a vector of initial
conditions at $t=0$ and $\exp(\cdot)$ denotes the matrix exponential
function.

How can we uncover network size, i.e. find how many dynamical variables
$N$ the system has if we measure the dynamics of only $n<N$ variables?
Without loss of generality, we observe the first $n$ components of
$\boldsymbol{x}(t)$ and all other $h=N-n$ state variables are hidden
from measurement. The time series of measured states $\boldsymbol{y}(t):=\left[x_{1}(t),x_{2}(t),\ldots,x_{n}(t)\right]^{\mathsf{T}}\in\mathbb{R}^{n}$
then satisfy the projection 
\begin{align}
\boldsymbol{y}(t)=\begin{bmatrix}I_{n} & 0\end{bmatrix}\boldsymbol{x}(t)=\begin{bmatrix}I_{n} & 0\end{bmatrix}\exp(At)\boldsymbol{x}(0),\label{eq:projection}
\end{align}
where $I_{n}$ is the $n\times n$ identity matrix and $0$ represents
the $n\times h$ matrix full of zeros. Thus we obtain the constraint
\begin{equation}
y_{i}(t)=\sum\limits _{j=1}^{N}\theta_{ij}(t)x_{j}(0)\label{eq:constraint}
\end{equation}
for every component $i\in\left\{ 1,2,\ldots,n\right\} $, where $\theta_{ij}(t)=\left[\exp\left(At\right)\right]{}_{ij}$
is some unknown, time-dependent function and $x_{j}(0)$ is the $j$th
component of the initial state, equally unknown for $j\in\{n+1,\ldots,N\}$.
Our central question is now: can we find $h=N-n$ despite these many
unknowns?

Rewriting the constraint \eqref{eq:constraint} in matrix form yields
\begin{equation}
\boldsymbol{y}^{(m)}(t)=\varTheta(t)\boldsymbol{x}^{(m)}(0),\label{eq:theta_vector}
\end{equation}
where $\varTheta(t)\in\mathbb{R}^{n\times N}$ and $\boldsymbol{y}^{(m)}(t)$
is the $m$-th observable trajectory at time $t$ generated from complete
initial conditions $\boldsymbol{x}^{(m)}(0)$, different for different
$m$. Considering $M$ different trajectories yields a system $Y(t)=\varTheta(t)X_{0},$
where $Y(t):=\left[\boldsymbol{y}^{(1)}(t),\boldsymbol{y}^{(2)}(t),\ldots,\boldsymbol{y}^{(M)}(t)\right]\in\mathbb{R}^{n\times M}$
is the matrix of known dynamical states at time $t$ and the matrix
$X_{0}:=\big[\boldsymbol{x}^{(1)}(0),\boldsymbol{x}^{(2)}(0),$ $\ldots,\boldsymbol{x}^{(M)}(0)\big]\in\mathbb{R}^{N\times M}$
collects different initial conditions. If these trajectories are sampled
at $k$ different time points $t_{1},\ldots,t_{k}$, for each trajectory
measured relative to its initial time, we group all values of $Y(t)$
evaluated up to time $t_{k}$ into a \emph{detection matrix} 
\begin{equation}
T_{(k,M)}=\Theta_{(k)}X_{0},\label{eq:CORE}
\end{equation}
where $T_{(k,M)}\left(t_{1},\ldots,t_{k}\right):=\left[Y(t_{1})^{\mathsf{T}},\ldots,Y(t_{k})^{\mathsf{T}}\right]^{\mathsf{T}}\in\mathbb{R}^{kn\times M}$
and $\Theta_{(k)}\left(t_{1},\ldots,t_{k}\right):=\big[\varTheta(t_{1})^{\mathsf{T}},$
$\ldots,\varTheta(t_{k})^{\mathsf{T}}\big]^{\mathsf{T}}\in\mathbb{R}^{kn\times N}$
\footnote{We remark that double transposition is required and that $T_{(k,M)}\left(t_{1},\ldots,t_{k}\right)\neq\left[Y(t_{1}),\ldots,Y(t_{k})\right]$}.
We note that here the lower indices $k,M$ refer to the size ($kn\times M$)
of the detection matrix, not to any element of a matrix.

Equation \eqref{eq:CORE} linearly relates the detection matrix $T_{(k,M)}$
assembled from the $M$ different time series sampled at $k$ different
times each, to unknown maps $\Theta_{(k)}$ encoding the dynamical
evolution (i.e. consequences of the flow of the system) and to the
initial conditions $X_{0}$ with also $(N-n)M$ unknown elements.
Despite little is known about $\Theta_{(k)}$ and $X_{0}$, the time
series merged into the linear system \eqref{eq:CORE} already provide
valuable information about the network size $N$. Specifically, 
\begin{equation}
\rank\left(T_{(k,M)}\right)\leq\min\left(\rank\left(\Theta_{(k)}\right),\rank\left(X_{0}\right)\right),\label{eq:rankInequality}
\end{equation}
and the rank of $T_{(k,M)}$ generically increases with increasing
the number $M$ of time series ($\rank(X_{0})=\min\left(N,M\right)$),
as well as with increasing the number of sampling points $k$ on each
of them, because the rank of $\Theta_{(k)}$ increases ($\rank(\Theta_{(k)})=\min\left(kn,N\right)$),
until the rank is maximal and equals $N$. Merging sufficiently many
time series, $M>N$, of sufficient length $k>N/n$ we obtain $\rank\left(\Theta_{(k)}\right)=\rank\left(X_{0}\right)=\rank\left(T_{(k,M)}\right)=N$.
At this point, adding more time series, i.e. increasing $M$, or extending
observations on each of them, i.e. increasing $k$, does not further
increase $\rank\left(T_{(k,M)}\right)$ so computing the rank of the
detection matrix $T_{(k,M)}$ assembled from time series of the subset
of the $n$ measured units yields the network size $N$ via \eqref{eq:CORE}.
Thus, 
\begin{equation}
\hat{h}=\rank\left(T_{(k,M)}\right)-n.\label{eq:hidden}
\end{equation}
is the estimated number of hidden units. Interestingly, there is no
principal lower bound on how small $n$ must be for this relation
to hold theoretically. In practice, measurement errors, noise and
limits in the detection matrix condition number \cite{stoer2013introduction}
limit feasible ratios $n/N$, see our analyses below.

\emph{Algorithm for detecting network size from time series data.}
One practical way of inferring network size through the rank inequality
\eqref{eq:rankInequality} is to numerically compute the ordered singular
values $\sigma=(\sigma_{1},\ldots,\sigma_{b})$ of $T_{(k,M)}$ such
that $\sigma_{1}\geq\sigma_{2}\geq...\geq\sigma_{b}$ , where $b=\min\{kn,M\}$
specifies the number of singular values, and to detect the largest
$\Delta_{\mathsf{max}}$ of the gaps 
\begin{equation}
\Delta{}_{j}=\log(\sigma_{j})-\log(\sigma_{j+1})\label{eq:logarithmicSingularValueGap}
\end{equation}
on the logarithmic scale. To safely detect the network size $N$ given
a known number $n$ of measured units from iteratively increasing
the number of measurements $M$ (see Fig.~\ref{fig:sketch}c), we
propose the following algorithm: 
\begin{enumerate}
\item Start, given the lower bound $n\leq N$, with a set of $M=n+1$ measurement
trajectories $\boldsymbol{y}^{(m)}(t)$, $m\in\{1,\dots,M\}$. 
\item Choose $k=\left\lceil \frac{M}{n}\right\rceil $ different
time instants $t_{\kappa}\in\{t_1,\ldots,t_{k}\}$ separated by $\Delta t=t_{\mathrm{tot}}/k$,
where $t_{\mathrm{tot}}$ is the total duration of each time series considered and $t_1$ its start time. 
\item Construct the detection matrix 
\begin{align}
T_{(k,M)}=\begin{bmatrix}\boldsymbol{y}^{(1)}(t_{1}) & \dots & \boldsymbol{y}^{(M)}(t_{1})\\
\vdots &  & \vdots\\
\boldsymbol{y}^{(1)}(t_{k}) & \dots & \boldsymbol{y}^{(M)}(t_{k})
\end{bmatrix}
\end{align}
from the measurements $\boldsymbol{y}^{(m)}(t)$ and compute its $b=\min\{kn,M\}=M$
singular values $\sigma\left(T_{(k,M)}\right)$. 
\item Compute logarithmic gaps $\Delta{}_{j}$ as in \eqref{eq:logarithmicSingularValueGap}. 
\item Save the largest gap $\widetilde{N}_{n}^{(M)}:=\max\{\Delta{}_{j}\}$,
where $j\geq n$ and $j\not\in\left\{ n,2n,\ldots\right\} \cup\{n+1,2n+1,...\},$
avoiding gaps at integer multiples of $n$. 
\item To robustly identify size also in case $N$ is such an integer multiple,
repeat steps 2--5 for $n-1,...,n-4$ measured units (thus ignoring
actually measured units) and take as the estimate 
\begin{equation}
\hat{N}^{(M)}:=\mathrm{median}\{\widetilde{N}_{n}^{(M)}\}.\label{eq:Nestimator}
\end{equation}
\item If $\hat{N}^{(M)}$ does not increase further, stop and define $\hat{N}:=\hat{N}^{(M)}$
as an estimate of network size; otherwise, repeat steps 2--6 with
one additional measurement, $M\rightarrow M+1$; 
\end{enumerate}
Here, step 2 ensures that finally, we will have $kn>N$ because $M>N$,
see the examples below.

\begin{figure}[ph]
\includegraphics[scale=0.6]{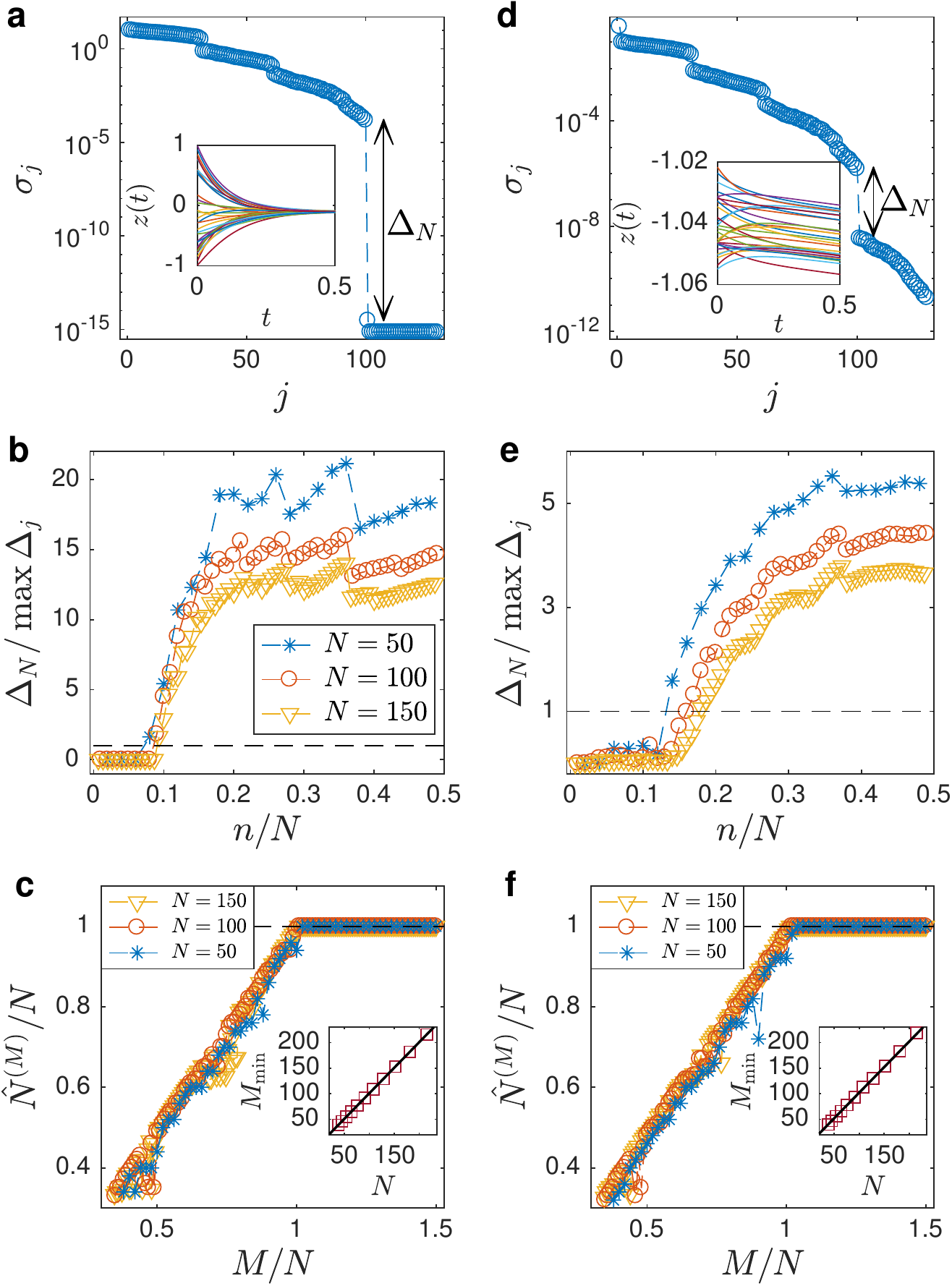} \caption{\textbf{Singular values of detection matrix yield network size. }\emph{(a)}
Singular values $\sigma_{j}$ of detection matrix $T_{(k,M)}$ displayed
on logarithmic scale for linear networks of $N=100$ diffusively coupled
units ($n=30$ measured). The largest gap $\Delta{}_{N}$ reveals
the size of the network. \textit{Inset: } Example trajectories $z_{i}(t).$
\emph{(b)} Size of $\Delta{}_{N}$ relative to largest $\Delta{}_{j}$
for $j<N$ rises above detection threshold at unity (horizontal dashed
line). Every data point corresponds to an average over 20 independent
random networks ($M=1.5N$). \emph{(c)} For increasing number of experiments
$M$, the inferred number $\hat{N}^{(M)}$ of units proportionally
increases until it stays constant at $\hat{N}^{(M)}=N$ once $M>N$.
\emph{Inset: } Minimum number $M_{\textnormal{min}}$ of experiments
to achieve $\hat{N}^{(M)}=N$ for networks of different sizes $N$
(red squares, $n=N/3$ measured units). All results well fit the prediction
$M_{\textnormal{min}}=N+1$ (solid line). \emph{(d)} Gaps revealing
network size for Kuramoto oscillators near a phase-locked state ($N=100$,
$n=30$ as before). \emph{(e)} As the theory uses linear approximations,
the strength of predictions is weaker for nonlinear systems, but the
relative gap size again clearly grows beyond unity (dashed line) with
increasing fraction $n/N$ of measured units. \emph{(f)} Even though
$\Delta_{N}$ is smaller compared to the linear dynamics, the same
method infers $N$ just as reliably. Here, the prediction is $M_{\textnormal{min}}=N+2$
as one measurement time series is used as a reference. We use homogeneous,
directed random graphs with in-degree $g=N/10$ (see Supplement).
\label{fig:gaps} }
\end{figure}

\emph{Performance of network size detection.} To test the predictive
power of our theory combined with the simple algorithm provided we
inferred the network size for five different classes of network dynamics:
(i) noiseless, diffusively coupled one-dimensional linear units collectively
converging to stable fixed points, (ii) phase-oscillator networks
close to periodic phase-locked states, systems of $N$ three-dimensional
coupled oscillatory units that exhibit (iii) regular periodic as well
as (iv) irregular chaotic collective dynamics, and (v) noisy, heterogeneous
systems with nonlinear dynamics. For settings (i) and (ii), we define
the class of diffusively coupled systems of single-variable units
via \eqref{eq:nonlinear} with $F_{i}(\mathbf{z})=\omega_{i}+\sum_{j=1}^{N}A_{ij}f\left(z_{j}-z_{i}\right),$
where $f:\mathbb{R}\rightarrow\mathbb{R}$ is a smooth function and
$\omega_{i}\in\mathbb{R}$ is a constant driving signal. We provide
all model and simulation details in the Supplement.

For the simplest setting of linear noiseless systems, we take $f(x)=x$
with stable fixed point $\boldsymbol{z}^{*}$ (Fig.~\ref{fig:gaps}a-c).
The estimated rank of the detection matrix \eqref{eq:CORE} indicated
by a pronounced gap in its singular value spectrum accurately predicts
network size (Fig.~\ref{fig:gaps}a) and is reliable already if only
about 10\% of the units are measured (Fig.~\ref{fig:gaps}b). Measuring
larger fractions $n/N$ of units rapidly further improves distinguishing
the largest gap $\Delta_{N}$ from other gaps $\Delta_{j}$. For nonlinearly
coupled systems of phase-oscillators ($f(x)=\sin(x)$, $\omega_{i}\in[-0.1,0.1]$),
performance is similarly high despite locally linear approximations
(Fig.~\ref{fig:gaps}d-f). We expected this similarity in performance,
because phase-locked states map to fixed points in a co-rotating frame
of reference and linearization of the sine function constitutes a
well-conditioned approximation for $|x|\ll\pi/2$.

\begin{figure}[t]
\includegraphics[scale=0.6]{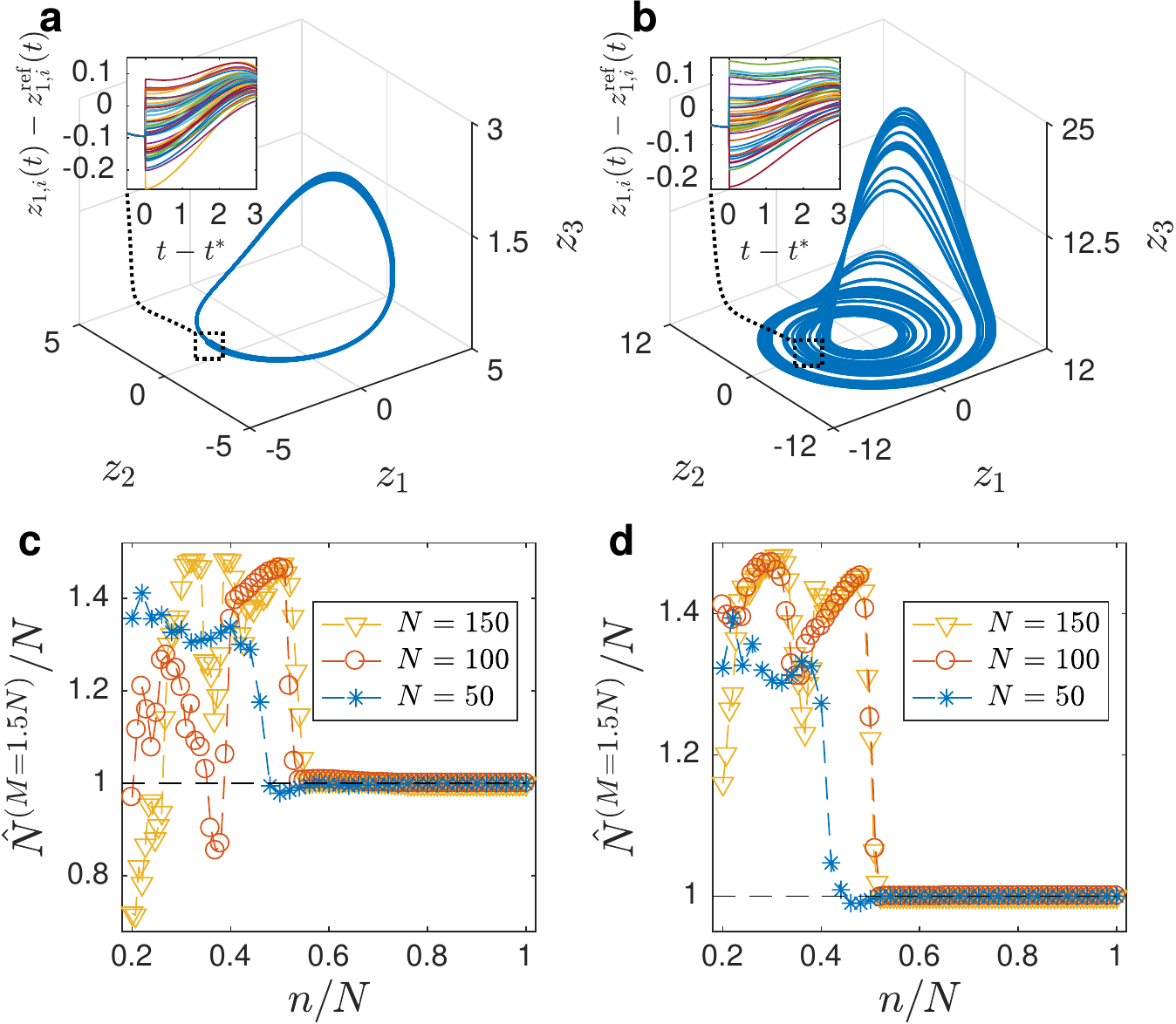} \caption{\textbf{Network size from complex transient dynamics.} Projection
of sample trajectories of one unit $i$ for \emph{(a)} periodic and
\emph{(b)} chaotic dynamical regimes. Each time, the system passes
a certain region on the attractor (highlighted by dashed square),
a random perturbation is applied to the components $z_{1,i}$ (insets).
\emph{(c,d)} Using deviations $\delta z_{i}^{(m)}(t)=z_{1,i}^{(m)}(t-t_{m}^{*})-z_{1,i}^{(1)}(t-t_{1}^{*})$
for each perturbation experiment $m$ to construct $T_{(k,M)}$ reveals
the correct system size $\hat{N}^{(M)}=N$, if a sufficient fraction
$n/N$ of units is measured. All data points averaged over 20 random
network realizations of $N=100$ units with degree ten, exhibiting
R\"ossler oscillatory dynamics, with state $\mathbf{z}_{i}(t)=(z_{1,i}(t),z_{2,i}(t),z_{3,i}(t))$,
and diffusive coupling between $\boldsymbol{z}_{2}$-components. In
the examples shown, the $z_{1}$-components of units $i$ are perturbed
and measured. Despite the coupling being in the $z_{2}$ components,
network size identification is accurate at $\hat{N}/N=1$.}
\label{fig:roessler} 
\end{figure}

\emph{Complex transient dynamics and biological networks.} The idea
introduced above is readily generalized to systems of higher-dimensional
units and more complex forms of collective dynamics, including in
principle arbitrary periodic or chaotic motion. Now consider that
$\boldsymbol{z}^{*}$ is not a fixed point of the dynamics \eqref{eq:nonlinear}
but \emph{any point in state space.} We locally approximate near $\boldsymbol{z}^{*}$
the nonlinear flow $\boldsymbol{\Phi}_{t}(\cdot )$ \cite{hale2012dynamics}
defined for all solutions $\boldsymbol{z}(t)$ of the original nonlinear
differential equation \eqref{eq:nonlinear} via $\boldsymbol{z}(t)=\boldsymbol{\Phi}_{t}(\boldsymbol{z}(0))$
from some initial conditions $\boldsymbol{z}(0)$. The difference
vector $\delta\boldsymbol{z}(t)=\boldsymbol{z}^{(1)}(t)\boldsymbol{-z}^{(2)}(t)$
of two close-by trajectories indexed $1$ and $2$ then satisfies
(see Supplement for a step-by-step derivation) 
\begin{align}
\delta\boldsymbol{z}(t) & \doteq D\boldsymbol{\Phi}_{t-t^{*}}\Big|_{\boldsymbol{z}^{*}}\delta\boldsymbol{z}(t^{*})\label{eq:generalFLOWcontraints}
\end{align}
where $D\boldsymbol{\Phi}_{t-t^{*}}\Big|_{\boldsymbol{z}^{*}}$ denotes
the Jacobian matrix of $\boldsymbol{\Phi}_{t-t^{*}}(\cdot)$ evaluated
at $\boldsymbol{z}^{*}$ and the symbol ``$\doteq$'' indicates
first order approximation in the components of $\delta\boldsymbol{z}(t^{*})$.
Employing a projection equivalent to \eqref{eq:projection} above,
we now take the time series of the measured units to be 
\begin{equation}
\boldsymbol{y}(t)=\begin{bmatrix}I_{n} & 0\end{bmatrix}\delta\boldsymbol{z}(t),\label{eq:FLOWprojection}
\end{equation}
the matrix generating the dynamics to have elements 
\begin{equation}
\theta_{ij}(t):=\left(D\boldsymbol{\Phi}_{t-t^{*}}\Big|_{\boldsymbol{z}^{*}}\right)_{ij}=\frac{\partial\Phi_{i,t-t^{*}}}{\partial z_{j}}\Big|_{\boldsymbol{z}^{*}}\label{eq:ThetaMatrixForFLOWs}
\end{equation}
and re-obtain \eqref{eq:constraint} for the difference variables.
We emphasize that the resulting equations are mathematically identical
to \eqref{eq:constraint} such that combining time series data as
before into a detection matrix yields the network size exploiting
the same principles and steps as above. In simulations, we consider
$\boldsymbol{z}^{(2)}(0)=\boldsymbol{z}^{*}$ for simplicity and thus
consider $t^{*}=0$ and positive times $t>0$. Figure \ref{fig:roessler}
illustrates successful network size identification for high-dimensional
periodic motion and for collective chaotic dynamics.

\begin{figure}[t]
\includegraphics[scale=0.6]{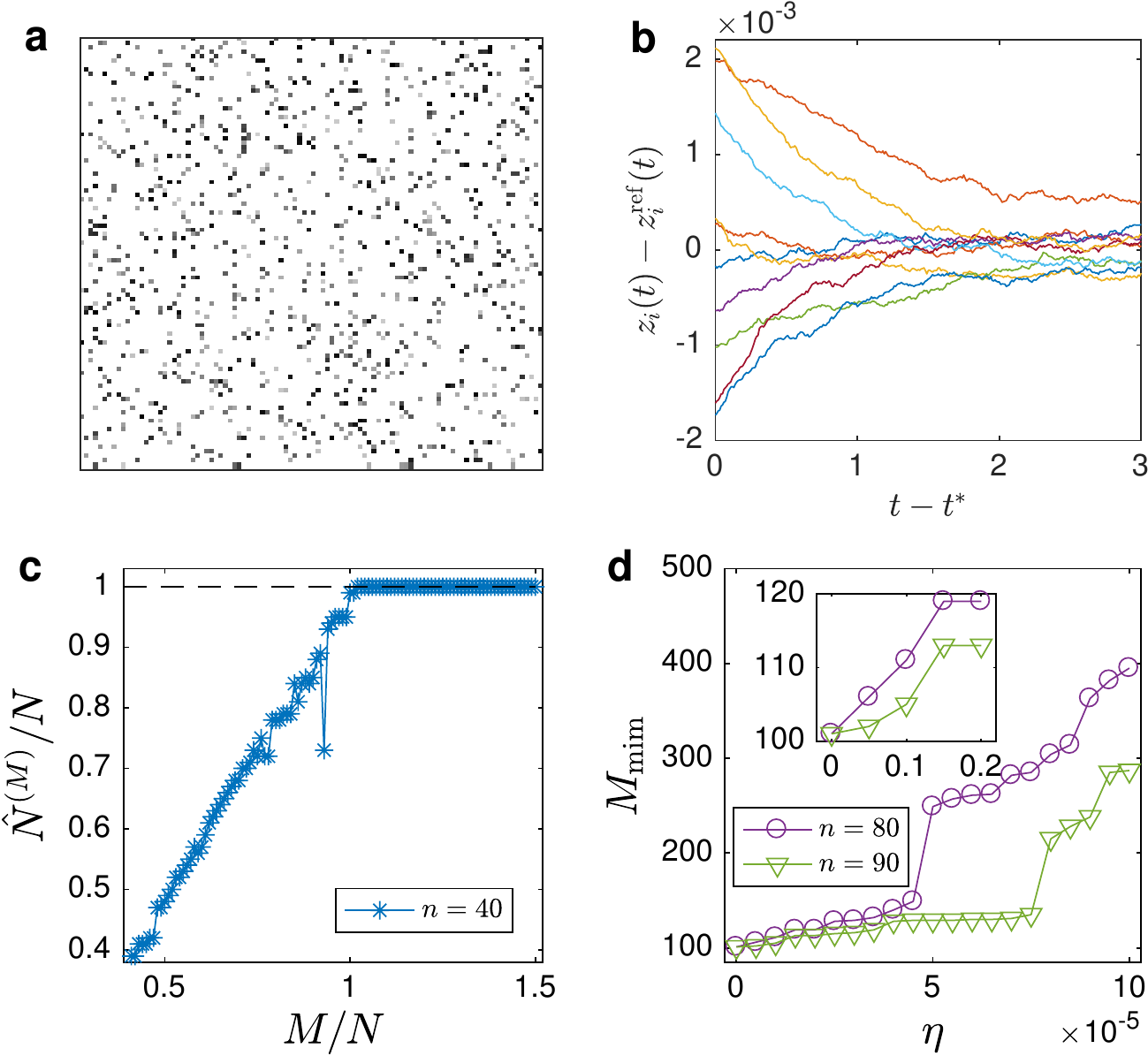} \caption{\textbf{Exact size detection in biological circuits simultaneously exhibiting
nonlinearities, heterogeneities, and noise.} \emph{(a)} Adjacency
matrix of a coupled Michaelis Menten kinetic network ($N=100$, link weights
in grayscale) and \emph{(b)} its collective noisy dynamics (units
of ten randomly selected units displayed, $\eta=10^{-4}$). As for
coupled periodic and chaotic systems, deviations $\delta z_{i}^{(m)}(t)=z_{i}^{(m)}(t-t_{m}^{*})-z_{i}^{(1)}(t-t_{1}^{*})$
are used for the reconstruction. \emph{(c)} Increasing the number
$M$ of measurements taken into account in the detection matrix reveals the network size once $M>N$ in the absence of noise.
\emph{(d)} The minimum number $M_{\mathrm{min}}$ of experiments required
to obtain an \emph{exact} size prediction $\hat{N}^{(M)}=N$ for $M\protect\geq M_{\text{min}}$,
in dependence of the noise level $\eta>0$.}

\label{fig:complex} 
\end{figure}

To illustrate applicability to biological circuits, we tested networks
displaying Michaelis Menten kinetics, a paradigmatic model of biochemical
reaction dynamics (see Figure \ref{fig:complex} and Supplementary Material). Intriguingly, exact size detection is feasible even in such systems simultaneously exhibiting nonlinearities, heterogeneities and noise. Most interestingly, detection may be exact despite noise. An increasing number of time
series taken into account still enables exact size identification,
$\hat{N}=N$. See also the Supplement for a systematic evaluation
of the influence of noise. 

\emph{Discussion and conclusions. }
In
summary, we proposed a theory for determining the network size from
time series data sampled from of a potentially small subset of perceptible
units. The novel perspective introduced shifts the problem of determining the exact number
of hidden units to the task of recording a sufficiently large number
of different dynamical trajectories of the perceptible units. It offers a generic tool for detecting the network size or, more generally, the number of independent dynamical variables of multidimensional coupled systems, from a fundamental theorem of linear algebra applied to linear constraints on a suitably constructed detection matrix. The main conditions for applicability are that (i) $M>N$ trials are experimentally feasible and that (ii) the sampling is such that two or more data points on a given trajectory are sufficiently close in state space for the dynamics obtained from local linearization to well approximate the real dynamics. 
While the time steps $t_{2}-t_1,\dots,t_{k}-t_{k-1}$ need to be the same in
each measurement, we emphasize that only very few such points are
needed if not too few nodes are recorded. Network size may be detectable from as few as $k=2$ sample points per trajectory if more than half of all units are perceptible. Moreover, even in modular networks where most perceptible units are located in one module, network size detection may work reliably (see also Supplemental Material at \url{https://doi.org/10.1103/PhysRevLett.122.158301}).

Compared to the state of the art, the conditions underlying network size identification can be considered mild, for at least two reasons. First, because so far only one or potentially a few individual hidden nodes are identifiable at all \cite{Su2012,Shen2014,Su2016} whereas our approach enables the identification of an extensive number of simultaneously hidden nodes. These may even be the majority of all nodes in the network. Second, because time series analysis methods of finding the attractor dimension (that constitutes a lower bound of and sometimes could equal the dimensionality of state space, and thus the number $N$ of active variables) require $M'\gg N$ data points and in addition are typically limited to moderate or even small $N$ of the order of ten or lower \cite{kantz2004nonlinear}. For example, to obtain faithful attractor dimensions that constitute lower bounds on $N$, as many as $M'>10^4$ data points may be required for systems with $N=3$ active variables \cite{pecora2007unified}, whereas our method requires $M'=kM$ data points with moderate or small $k\geq 2$ and $M$ just slightly larger than $N$.

We tested the proposed theory employing direct numerical simulations of abstract model systems and generic biological circuit models simultaneously exhibiting three obstacles that may limit network size detection. We find that for successful detection, collective network dynamics may be non-stationary, it may be close to fixed points or more complex such as  periodic or possibly aperiodic chaotic or noisy motion. A case study of generic model biological circuits equally reveals network size despite simultaneously exhibiting nonlinearities, heterogeneities and noise. Across these settings, size detection may be \emph{exact}. The applicability may be limited under conditions where noise strongly dominates the dynamics, only a small fraction of units are perceptible or perceptible units in a modular network are located all in one module. Example studies (see Supplemental Figure S2 for an illustration) suggest that even in the latter extreme setting, although the exact size is not revealed any more, the estimate is still of the same order of magnitude.

A related challenge is network observability \cite{Liu2460,TANG2014184,PhysRevX.5.011005,parlitz2016estimating},
that is to identify a sufficient set of units such that measuring
these units' states reveals the collective state of the entire network.
In contrast, our work aims at identifying the number of units in a
network, not their states. It is thus conceptually different and exhibits
much weaker requirements.

Previous approaches to detect hidden nodes
are capable of detecting a single hidden node in an otherwise completely
perceptible network: Some \cite{Hamilton2017} employ nonlinear Kalman
filters to fit the parameters of a given model and use the covariance
matrix of the fitting error; others first approximate the dynamics
via differential equations and then determine the existence and location
of the hidden unit through heuristic methods \cite{Su2012,Shen2014,Su2016}.
Our theory instead reliably captures many hidden units, is data-driven,
relies on sampled time series and thereby requires
no model \emph{a priori}. 
Furthermore,
it provides a mechanistic perspective that not only determines the
existence but also reveals the exact number of hidden units. 
It may thus also complement embedding methods for determining attractor dimension
\cite{kantz2004nonlinear} that identify the number of active variables
from stationary time series, thereby opening
up a way to broaden insights about the collective dynamics of multi-dimensional complex systems. \cite{parlitz2016estimating}.

\begin{acknowledgments}
We gratefully acknowledge support from the Ministry for Science and
Culture of the German Federal State of Lower Saxony (grant no. ZN3045,
nieders. Vorab to H.H. and J.P.), the German Federal Ministry of Education
and Research (BMBF grants no. 03SF0472F and 03EK3055F to M.T.), the Deutsche Forschungsgemeinschaft (DFG, German Research Foundation) with a grant towards the Cluster of Excellence Center for Advancing Electronics Dresden (cfaed) and under Germany's Excellence Strategy - EXC-2068 - 390729961 - Cluster of Excellence Physics of Life at TU Dresden.
\end{acknowledgments}


%

\end{document}